\documentclass[aps,pre,twocolumn,endfloats]{revtex4}

\usepackage{psfrag}
\usepackage{graphicx}
\usepackage{bm}

\newcommand{\cz}[0]{[\makebox[0.2cm]{ }]}
\newcommand{\Tau}[0]{{\cal T}}
      
\begin{document}       

\bibliographystyle{plain}
       
%\draft        

\title{The interplay between chemical reactions and transport in
structured spaces}

\author{Zoran Konkoli}
\email{zorank@fy.chalmers.se}
\affiliation{       
   Department of Applied Physics,\\      
         Chalmers University of Technology and G\"oteborg University, \\  
         SE-412 96 G\"oteborg, Sweden
} 
\date{\today}

\begin{abstract}      
There are many instances in nature where geometry of the system that
sustains chemical reactions is structured and the living cell is a
typical example. There is a high degree of compartmentalization in the
living cell where various compartments sustain different chemical
reactions and transport reactants among themselves.  In order to avoid
storage facilities reactants are routed in orderly fashion between
various places in the cell interior with large degree of
synchronization. It is exactly such situation that is investigated
here. The main motivation behind this study is to understand interplay
between reactions and transport in a geometries that are not
compact. Typical examples of compact geometries are box, sphere,
etc. On the other hand, a network made of containers
$C_1,C_2,\ldots,C_N$ and tubes is a typical example of space that is
structured, and such non compact space is main focus on this
study. The whole space is divided into a two regions. First, in
containers particles react with rate $\lambda$. Second, tubes
connecting containers allow for exchange of chemicals with transport
rate $D$.  In such a way the two most important processes are isolated
in the problem, reactions and transport. By varying topology of such
network and details of chemical reactions it is possible to gain some
understanding of interplay between chemical reactions and transport in
structured spaces. It assumed that the number of reactants in the
system is so small that kinetics is noise dominated. Two methods for
solving corresponding master equation are discussed. The computer
simulation is easy to implement, but leads to the results that are not
that accurate. In here, a method is presented that can be used to
calculate average, variance, and higher moments of the time reaction
needs to finish. The method relies on a matrix representation of the
master equation and is in principle exact. It works for an arbitrary
reaction scheme and network topology.  A number of different chemical
reactions were studied and their performance compared in a various
ways. Reactions are grouped into two ensembles, Reaction on a Fixed
Geometry Ensemble (ROGE) and Geometry on a Fixed Reaction Ensemble
(GORE). The ROGE and GORE are used to classify reactions in order to
gain some understanding of which types of chemical reactions draw most
benefit from the structured spaces. Most important findings are as
follows. (i) There is a large number of reactions that run faster in a
network like geometry. Such reactions contain antagonistic catalytic
influences in the intermediate stages of a reaction scheme that is
best dealt with in a network like structure. (ii) Antagonistic catalytic
influences are hard to identify since they are strongly connected to
the pattern of injected molecules (inject pattern) and depend on the
choice of molecules that have to be synthesized at the end (task
pattern). (iii) The reaction time depends strongly on the details of the
inject and task patterns.
\end{abstract}


\maketitle

%\begin{multicols}{2}       
       
%\narrowtext       
       
\section{Introduction}

The goal of the present study is to impact some progress in
understanding interplay between reactions and transport in structured
spaces.  There are many questions one could ask. This study focuses on
one: How strongly the geometry of the system (e.g., topology and shape
of the boundaries) influences the particular reaction schemes it
harbors? This question can be asked in two ways.  First, assuming that
spatial structure is fixed, which reaction scheme would draw most
benefit from it, for example, in terms of speed (smaller execution
time) or better timing (reduction of noise level)?  Second, given the
chemical reaction, what is the topology that is mostly suited for it?
In order words, in which ways should one alter geometry of the system
(without altering reactants or influencing reaction mechanisms in any
other way) in order to speed up chemical reaction encapsulated inside?
Answers to these questions are relevant to a number of topics ranging
from understanding of physio-chemical processes in the living cell,
biological evolution, towards chemical engineering and biotechnology.

Taking the living cell as an example. The living cell exhibits large
degree of spatial organization. Various compartments sustain different
chemical reactions and transport reactants among themselves, with a
high degree of spatial and temporal synchronization.  Reactants and
products of reaction are transported where they are needed, exactly at
the right place and at the right time. There is no need of excessive
storage of molecules. For reviews on these topic
see~\cite{Nelsestuen,Kuthan}, and
refs.~\cite{Ganti,SSFA,MH,SMH1,SMH2,SZMH,SMH3,WQL,SQ,QQ} for other
interesting work. The topics related to temporal and spatial
synchronization of the cellular machinery are still hotly debated and
present work might offer some understanding of these
phenomena. Furthermore, thinking from evolutionary perspective, one
wonders what is that cells try to optimize by adopting highly
structured geometry?

Evolutionary process occurred in two stages: the chemical
evolution~\cite{Calvin}, which lead to emergence of organic molecules
necessary for synthesis of proteins and anything alive, followed by
biological evolution~\cite{Volkenstein}, a process which started after
emergence of the first living cell.  Fairly little has been done in
understanding the role of geometry and spatial organization played in
the evolution process.  There is already some pioneering work in this
area~\cite{Langton,Furusawa,Vespalcova,Ono,Boer}. Present study could
impact some progress in this direction.

Besides the topics discussed, the question of interplay between
geometry and reaction scheme is relevant to a number of disciplines. The
chemical engineering and biotechnology are typical examples. In
general, there is a tendency to move away from bulk situation where
volumes are large and reactants many, towards more exotic regime where
reaction volumes are small and structured and number of reactants
relatively few. The interesting experimental work on these topics can
be found in refs.~\cite{Owe1,Owe2,Owe3,Owe4} and
\cite{Ross1,Ross2,Ross3}. The first set of references deals with
reactions in liposome networks, and second set addresses the situation
where a number of perfectly stirred reactors is connected by
tubes. The exchange of chemicals through tubes is driven by pumps. The
work done in here is relevant to both of these studies.

The particular model proposed here is influenced by several lines of
research. The setup of the reaction schemes is inspired by the work
done on prebiotic evolution, genome-based evolution and random
reaction
networks~\cite{Bagley1,Bagley2,Farmer,Kauffman,Stadler,Schuster,Slanina,Jain},
and diffusion-controlled reactions in bulk
phase~\cite{Kotomin1,Kotomin2,CompChemKin,Kotomin3,Mikhailov1,Ovchinnikov}
and in restricted
geometries~\cite{McQuarrie1,Clifford1,Khairutdinov1,KKO}.  The
geometrical setup of the model is motivated by the experimental
studies given in ~\cite{Owe1,Owe2,Owe3,Owe4,Ross1,Ross2,Ross3} and
theoretical study of reaction-diffusion neuron and enzymatic
neuron~\cite{Akingbehin,AkinCon,Conrad4,KirCon1,KirCon2,KamCon1,KamCon2}.

\begin{figure}       
\psfrag{AA}{(a)}
\psfrag{BB}{(b)}
\psfrag{CC}{(c)}
\psfrag{C1}{$C_1$}
\psfrag{C2}{$C_2$}
\psfrag{C3}{$C_3$}
\psfrag{C4}{$C_4$}
\psfrag{C5}{$C_5$}
\psfrag{L15}{$l_{15}$}
\psfrag{L25}{$l_{25}$}
\psfrag{L35}{$l_{35}$}
\psfrag{L45}{$l_{45}$}
%%%%%%%%%%%%%%%%%%%%%%%%%%%%%%%%%%%%%%%
\includegraphics[width=8cm]{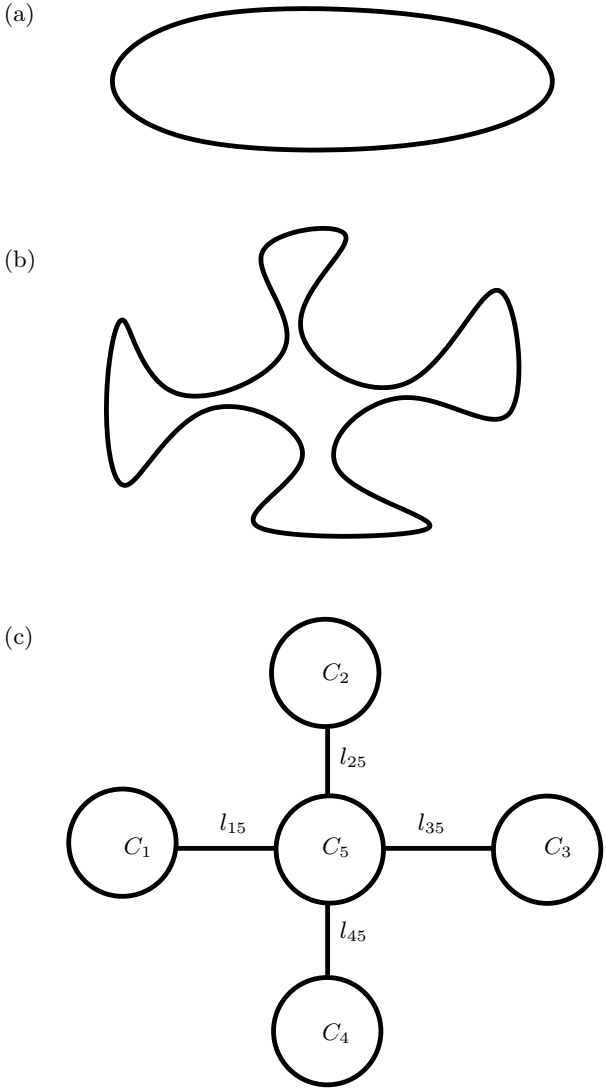}
%%%%%%%%%%%%%%%%%%%%%%%%%%%%%%%%%%%%%%%%
\caption{ A guidance how to think about the problem. Panel (a):
non-structured (compact) space. Panel (b): structured space obtained
by deforming geometry depicted in panel (a).  The goal is to
understand how reaction dynamics alters when geometry changes from (a)
to (b). The panel (c), representing a network of containers and
tubes, is used to achieve this goal.  For example, to capture the most
important geometrical features of the structure in panel (b) one needs
five containers $C_1,\ldots,C_5$ and four tubes with lengths $l_{15}$,
$l_{25}$, $l_{35}$ and $l_{45}$. The two parameters are used to define
the dynamics in the network (c). The intra-container reaction rate
$\lambda$, and inter-container transport rate $D$ (provided containers
are connected). When $D\sim\lambda$ the reaction dynamics in panels (b)
and (c) should exhibit some similarities. For $D\gg\lambda$ one
expects the same for the structures (a) and (c).}
\label{network}       
\end{figure}

The situation considered here is radically different from the bulk
studies. The goal is to mimic particular aspect of the cell
environment where reactions happen in the specific regions of space
and reactants move among these regions.  The simplest way of achieving
such a setup is to consider chemical reactions in containers that are
connected by tubes. Figure~\ref{network} shows one possible
example. Reactants can move from container to another along tubes.
Thus, the compartmentalization is build into the model in the simplest
possible way. Furthermore, it is assumed that containers are small in
size so that the number of reactants in each container is also
small. Such setup inevitably leads to fluctuations, and chemical
kinetics becomes noisy, as is usual the case in the living cell.

For simplicity reasons, the details, both of the chemical reactions and
transport, are neglected to a large extent. It is assumed that reactants
are point like objects with no structure. What is studied here can be
classified as artificial chemistry~\cite{Dittrich}. The two most
important time scales are traced in the model. The reaction rate
$\lambda$ describes how fast molecules react in containers.  The
transport rate $D$ governs exchange of chemicals among containers. It
is assumed that diffusion is the dominant mechanism of particle
transport and this is reasonably good approximation for the living
cell.~\cite{Nelsestuen,Kuthan} Naturally, there are exceptions to this
rule, e.g. transport of molecules by kinesin along microtubule in the
cell, but such cases are not going to be considered here. For
discussion of reaction-diffusion systems in biology please see
ref.~\cite{Hess}.

How to compare chemical reactions in a compact and structure
geometries depicted in Fig.~\ref{network} [panels (a) and (b)]?
Naturally, the most obvious way would be to run chemical dynamics
directly on structures depicted in panels (a) and (b) and try to trace
differences. However, it will be shown that this can also be done by
studying kinetics in network like geometry depicted in panel (c), but
such strategy has to be implemented carefully.  For example, it is
easy to see that panel (c) in Fig.~\ref{network} is a very rough
representation of structured geometry in panel (b). However, there is
no apparent similarity between structures shown in panels (a) and
(c). Nevertheless, the network structure in (c) contains (a) and (b)
as special cases. For example, when $D\sim\lambda$ the reaction
dynamics occurring in the network depicted in panel (c) should exhibit
roughly the same features as the reaction dynamics running in the
structured geometry depicted in panel (b). Furthermore, when transport
rate dominates all other processes in the system, $D\gg\lambda$, same
should hold for (a) and (c). Also, in such a case, $D$ should be much
larger than any other reaction rate that might enter due to the
effects of catalysis. Thus, in order to gain some understanding of
differences between (a) and (b) one simply has to study variations of
a reaction dynamics in the network like structure (c) as transport
rate changes from $D\sim\lambda$ towards $D\gg\lambda$.

The type of the analysis used is largely static. To understand in
which ways particular shape of the reaction volume influences chemical
kinetics (and vise versa) a large number of chemical reactions will be
analyzed and their performance compared. The set of reactions and
geometries chosen form an ensemble ${\cal
A}=\{o_\omega,\omega=1,\ldots,E\}$ where elements in ensemble are
pairs $o_\omega=({\rm reaction},{\rm geometry})$. The element $o_\omega$
will be referred to as an {\em organism}.

Two ensembles are distinguished. In the first case, the reaction
scheme is kept fixed, while the geometry of the network is allowed to
change.  This type of ensemble will be referred to as the geometry on
the (fixed) reaction ensemble (GORE). The geometries are different and
they are sampled by changing size of the containers and length of the
tubes. In the second case the geometry of the network is kept fixed,
while the reaction scheme is subject to a change. This type of the
ensemble will be referred to as the reaction on the (fixed) geometry
ensemble (ROGE). In this work most of the attention is on the ROGE
ensemble.

What is the measure of the good performance for a chemical reaction?  In
here, focus will be on the time related issues such as the length of
catalytic cycles.  The importance of the timing in the intra cellular process
has been discussed by Volkenstein \cite{Volkenstein}: `The most
important role in ontogeny is played by time factors, i.e. the time
relations in the synthesis of various proteins and the formation of
various cells and tissues. Even small changes in the timetable lead to
considerable morphological alternations...'. The setup presented here
allows for consideration of other performance criteria but these are
omitted due to the simplicity reasons.

The time needed for a reaction to finish, $\Tau$, is a stochastic
variable and for a given reaction scheme and topology it will have
well defined distribution function $\Phi(t)$: the probability that
reaction finished within the time interval $[t,t+dt]$ is given by
$P(t<\Tau<t+dt)=\Phi(t)dt$. All information about reaction timing is
hidden in the distribution function $\Phi(t)$. In practise, it is very
hard to find $\Phi(t)$ for general reaction scheme and geometry. Also,
amount of information contained in $\Phi(t)$ is too large. It is more
fruitful to characterize $\Phi(t)$ in terms of few variables and in
this work we use two: the average time needed for reaction to complete
$\tau=\int_0^\infty t\Phi(t)dt$ and the variance $\sigma$ where
$\sigma^2=\int_0^\infty(t-\tau)^2\Phi(t)dt$.  In such a way it is
possible to asses how {\em fast} reaction happens and how {\em noisy}
it is.

Once performance criteria are quantified one faces optimization
problem of finding the best performer in a given ensemble ${\cal
A}$. In principle, the best performer will be the organism that
exhibits smallest value for $\tau$. Two strategies are used to find
such organism. First, one simply generates full ensemble, measures
performance of each element, and sorts performance at the end.
However, when number of elements in such ensemble is large another
optimization method has to be used. For example, when there is a need
for that, one can use a method described in
refs.~\cite{Akingbehin,AkinCon,Conrad4,KirCon1,KirCon2,KamCon1,KamCon2}.
This optimization method mimics process of evolution and belongs to
the class of genetic optimization algorithms.  The terms evolution and
optimization will be used interchangeably through out the text.

The paper is organized as follows. In section \ref{sec:model} the
reaction-diffusion model is defined in detail, and in
section~\ref{sec:dynamics} overview of the methods for solving the master
equation is given. The method of doing a computer simulation is reviewed
first. The method of finding moments of the time length of the catalytic
cycles is discussed in detail in the next
section~\ref{sec:moment_method}. Few simple reactions are discussed in
section~\ref{sec:ROGE1} where it is shown how to compare vastly
different reaction schemes. Section~\ref{sec:ROGE2} contains analysis
of ROGE ensemble made from two container network, two particle types,
all possible reactions, and full set of inject-task
patterns. Conclusions and outlook are given in
section~\ref{sec:discussion}.

\section{The Model}
\label{sec:model}

The reaction-diffusion model is defined as follows. Consider the set
of containers $C_i$ connected in a particular way by tubes with
lengths $l_{i,j}$; $i,j=1,\ldots,N$. It is possible that some
containers are not connected. Example of such structure is given in
Fig. 1. The containers harbor molecules $X_\alpha$;
$\alpha=1,2,\ldots,M$.  Molecules are allowed to react only when in
the same container (provided there is a reaction they participate
in). Molecule $X_\alpha$ moves from the container $C_i$ to the
container $C_j$ with a rate $D^{\alpha\alpha}_{ij}$. Please note that
the expression for the transport rate is diagonal in the particle
index $\alpha$ (there is no change of particles while being
transported). This assumption could be easily relaxed. For simplicity
reasons it is assumed that $D^{\alpha\alpha}_{ij}=f(l_{ij})$ if the
link between container $i$ and $j$ exists, otherwise
$D^{\alpha\alpha}_{ij}=0$. Links are non-directional and transport
rates are same for all particles,
i.e. $D^{\alpha\alpha}_{ij}=D_{ij}=D_{ji}$. Please note that the
regular lattice is a special case of this very general model.
Structures like these can be found in the interior of the living
cells.  Mapping is not exact, but there is a rough similarity.

By assumption, the reactions only occur in the containers and not in
the tubes, and reactants mix well in the containers. Tubes serve only
as connectors of regions where reactions happen. With assumptions at
hand, to describe system at any time instant, it is sufficient to track
number of particles in each container. This greatly simplifies
calculations. Conformation of the system ${\bm c}$ is specified as
a occupancy of the containers,
\begin{equation}
  {\bm c} = ({\bm n}_1,...,{\bm n}_i,...,{\bm n}_N)
  \label{vecc}
\end{equation} 
where vectors ${\bm n}_i$ $i=1,...,N$ describe particle content of
each container,
\begin{equation}
  {\bm n}_i = (n_{1,i},...,n_{\alpha,i},...,n_{M,i})
  \label{vecn}
\end{equation}
with $n_{\alpha,i}=0,1,2,...,\infty$ for $\alpha=1,...,M$ and $i=1,...,N$.
System makes random transitions between various configurations. A
configuration of the system, ${\bm c}$, changes when either reaction or
transport occur.

For computational convenience, the very simple type of a reaction
scheme will be considered,
\begin{equation}
  X_\alpha \stackrel{\pm X_\gamma}{\rightarrow} X_\beta 
  \label{reaction}
\end{equation}
where $X_\gamma$ is understood to influence conversion of $X_\alpha$
to $X_\beta$ either as catalyst ($+$) or suppressor ($-$).  All
reactions that are allowed are assumed to have same reaction rate
$\lambda$.  The reaction graph is specified by the reactivity matrix
${\bm \Lambda}$.  If reaction $X_\alpha\rightarrow X_\beta$ is allowed
$\Lambda_{\alpha,\beta}=1$ and equals zero otherwise. The reaction
graph is directed and matrix ${\bm \Lambda}$ is not necessarily symmetric.

The effect of catalyst or suppressor are modeled as follows. The bare
reaction rate $\lambda$ may be modified by the presence of other
reactants in the container $C_i$.  It is assumed that each reaction
can have at most one catalyst or suppressor. On the other hand, it is
possible that one particle is catalyst (or suppressor) for more than
one reaction.  For programming purposes it is sufficient to use array
$K_{\alpha,\beta}=\pm\gamma$ if $X_\gamma$ is catalyst ($+$) or
suppressor ($-$) for $X_\alpha\rightarrow X_\beta$ reaction.
$K_{\alpha,\beta}=0$ indicates that reaction does not have catalyst
nor suppressor.  Also, it is assumed that the effects of the catalysis are
strongly enhanced. If there is some catalysis going on, it
completely dominates reaction scheme. Using these assumptions the
reaction rate for the reaction $X_\alpha\rightarrow X_\beta$ in the
container $C_i$ is given by
\begin{equation}
  \lambda^i_{\alpha,\beta}({\bm n}_i) = \lambda\Lambda_{\alpha,\beta}
  \left\{ 
    \begin{array}{ll}
       1  & \kappa=0 \\
       \xi  & \kappa>0, \  n_{\kappa,i} > \delta_{\kappa,\alpha} \\
       \frac{1}{\xi} & \kappa<0,\ n_{|\kappa |,i} > \delta_{|\kappa |,\alpha}
    \end{array}
  \right.
  \label{catalysis}
\end{equation}
$\delta_{\alpha,\beta}$ is a Kronecker delta-function, $\xi>1$ denotes
catalysis enhancement factor, and $\kappa=K_{\alpha,\beta}$.
Equation~(\ref{catalysis}) is self explanatory for the exception of
possibly one term that we proceed to discuss. The condition
$n_{\kappa,i}>\delta_{\kappa,\alpha}$ ensures that there is no self
catalysis for the reaction of the type
\begin{equation}
  X_\alpha\stackrel{X_\alpha}{\rightarrow}X_\beta
  \label{selfcat}
\end{equation}
when there is only one $X_\alpha$ present in the container. One needs at
least two $X_\alpha$ in order to fell catalytic influence of
$X_\alpha$ on the reaction given in (\ref{selfcat}).

\begin{figure}  
\psfrag{X1}{$X_1$}     
\psfrag{X2}{$X_2$}     
\psfrag{Xomega}{$X_\omega$}     
\psfrag{Xp1}{$X_1'$}     
\psfrag{Xp2}{$X_2'$}     
\psfrag{Xpeta}{$X_\eta'$}     
%%%%%%%%%%%%%%%%%%%%%%%%%%%%%%%%%%%%%%%
%\epsfxsize=8cm       
%\epsfbox{diagram.eps}
\includegraphics[width=8cm]{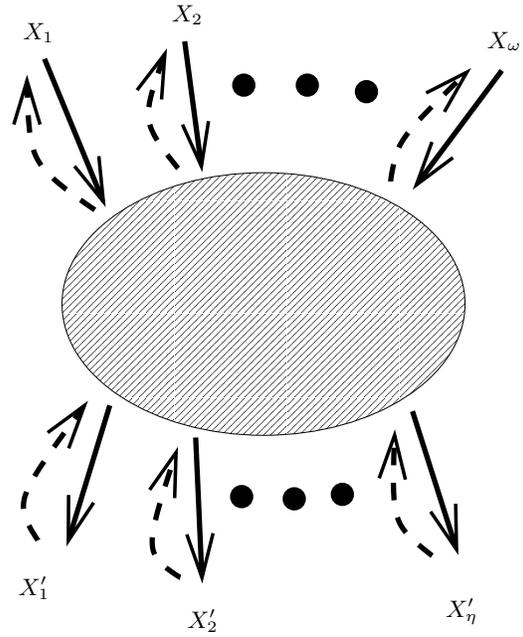}
%%%%%%%%%%%%%%%%%%%%%%%%%%%%%%%%%%%%%%%%
\caption{The most general form of a reaction graph considered.
Species $X_1,X_2,\ldots,X_\omega$ describe an {\em inject}
pattern, a set of the molecules that are inserted into the system at
$t=0$. This set is arranged into the vector $\bm\iota$.  Molecules
denoted by $X'_1,X_2',\ldots,X_\eta'$ are the ones that have to be
synthesized and are arranged into the vector $\bm\pi$. These are
referred to as a {\em task} pattern.  The shaded area in the middle
denotes intermediate reaction steps. Dashed lines with arrows are
allowed and result in appearance of loops in the reaction scheme.  The
speed of reaction is calculated by tracking instances when particles
in the set $\bm\pi$ appear for the first time. The corresponding times
$\Tau_1,\Tau_2,\ldots,\Tau_\eta$ are arranged into the vector
${\bm\Tau}$. See section \ref{sec:model} for more details.}
\label{diagram1}       
\end{figure}

The particular type of the reaction schemes considered here
(\ref{reaction}) is inspired by chemical processes in the living
cell. It is often that by action of enzymes certain input set of
chemicals is converted into output molecules by series of intermediate
reactions. It is exactly this aspect of cellular machinery which this
scheme tries to capture in the most simple way. The most general
reaction graph one can consider is shown in Fig.~\ref{diagram1}. The
graph indicates that cell machinery converts molecules
$X_1,...,X_\omega$ into molecules $X'_1,X'_2,...,X'_\eta$. The shaded
area in the middle denotes intermediate reaction steps that involve
arbitrary set of reactions between molecules already shown in the
graph. An additional particle types may appear in the shaded region.
It is possible to define the speed of a reaction as the time needed for the
predetermined set of output molecules $X'_1,X'_2,...,X'_\eta$ to
appear for the first time, provided only the input molecules were
present initially in the system. How this idea is implemented is shown
bellow.

In addition to reaction scheme given in Eq.~(\ref{reaction}) two
quantities are specified. First, at $t=0$ a certain number of particles is
injected into the system in various containers. The list of the particles
injected is specified by the vector 
\begin{equation}
  {\bm \iota} = (n_{1,1}^{(0)},\ldots,n_{\alpha,i}^{(0)},\ldots,n_{M,N}^{(0)})
  \label{inject}
\end{equation}
In the course of time the injected set of particles will
be transformed into something else. Second, the vector
\begin{equation}
  {\bm \pi} = (n_{1,1}^{*},\ldots,n_{\alpha,i}^{*},\ldots,n_{M,N}^{*})
  \label{task}
\end{equation}
specifies tasks that have to be achieved. For example,
$n_{\alpha,i}^*>0$ indicates that goal is to synthesize
$n_{\alpha,i}^*$ molecules of the type $X_\alpha$ in the container
$C_i$. On the other hand, $n_{\alpha,i}^*=0$ indicates that the number of
particles $X_\alpha$ in the container $C_i$ is not traced.  The maximum
number of tasks is given by $N\times M$, though not all tasks need to
be monitored. Also, the cases where task is achieved trivially at
$t=0$ are forbidden and additional restriction is set upon components
of ${\bm \iota}$ and ${\bm \pi}$: If task is traced one has
$n_{\alpha,i}^*>0$. In such case, the condition
$n_{\alpha,i}^*>n_{\alpha,i}^{(0)}$ must hold.

To make a notation easier, it is convenient to eliminate the elements from
${\bm\iota}$ and ${\bm\pi}$ that are zero and write
\begin{equation}
  {\bm\iota} = (\iota_1,\iota_2,\ldots,\iota_\omega), \ \ \ 
  {\bm\pi} = (\pi_1,\pi_2,\ldots,\pi_\eta)
  \label{task1}
\end{equation}
where ${\bm\iota}$, and ${\bm\pi}$, only contain
list of molecules injected, and tasks that are actually monitored. It is clear that $\omega,\eta\le M\times N$.

Every time a certain task is accomplished the time when this happens
is stored. These times are arranged in the vector 
\begin{equation}
  {\bm \Tau}=(\Tau_1,\Tau_2,\ldots,\Tau_\eta)
  \label{tau}
\end{equation}
and there is a one to one correspondence between the elements of
$\bm\pi$ and $\bm\Tau$. Once the task is achieved molecules that were
used to accomplish it are removed from the system.  This consideration
is motivated by the character of the real processes in the living
cell.  If certain number of molecules are needed at specific place,
once arriving there, these molecules will be consumed by other
biochemical processes in the living cell.

The vector ${\bm \Tau}$ is a stochastic variable and can be described in
terms of the distribution function
$\Phi(t_1,t_2,\ldots,t_n;\iota,\pi)$. In practise, it is very hard to
obtain full distribution function and it is more convenient to use
first two moments, the average
\begin{equation}
  {\bm \tau}=(\tau_1,...,\tau_k,...,\tau_n) 
  \label{avtau}
\end{equation}
and the variance 
\begin{equation}
  {\bm \sigma}=(\sigma_{1,1},...,\sigma_{\alpha,i},...,\sigma_{M,N})
  \label{sigma}
\end{equation}
In addition, one could also include moments of the type
$\langle\tau_i\tau_j\rangle$ $i\ne j$ but these will be not
considered.

The quadruple consisting of particular reaction scheme ($\lambda$,
$\xi$, ${\bm \Lambda}$ and ${\bm K}$), network geometry ($l_{i,j}$
$i,j=1,\ldots,N$), inject pattern ${\bm \iota}$, and list of tasks
monitored ${\bm \pi}$, will be referred to as an {\em~organism}. The
organism can be seen as the entity that has to transform a certain
number of chemicals into a set of molecules that have to be
synthesized at certain places, utilizing available reaction scheme and
geometry. In the following sections various organisms will be
classified according to the criteria how fast they achieve certain
list of tasks (see. Eq.~\ref{avtau}). In principle, it is possible to
make classification with regard to how noisy performance is (see
Eq.~\ref{sigma}), but this type of classification is left for future
studies.

\section{The dynamics}
\label{sec:dynamics}

The dynamics of the system defined in the previous section is stochastic
and from the rules discussed one can derive a master equation that
describes the time evolution of the occupation probabilities $p({\bm c},t)$,
\begin{equation}
  \dot p({\bm c},t)=  
     \sum_{{\bm c}'}R_{{\bm c},{\bm c}'} p({\bm c}',t) -
     \sum_{{\bm c}'}R_{{\bm c}'{\bm c}} p({\bm c},t)
  \label{MEQ}
\end{equation}
where here and in the following dot over symbol denotes time
derivative. The reaction rates $R_{{\bm c}'{\bm c}}$ describing
transition ${\bm c}\rightarrow{\bm c}'$ can be easily calculated from
the definition of the model.  In general, it is very hard to solve the
equation~(\ref{MEQ}). In here, two strategies are use to solve
it. First strategy is based on a simulation method. This is a straight
forward approach. In the second instance a set of equations is derived
that specifies first, second, and possibly higher moments of the
$\Phi(t_1,\ldots,t_n;\iota,\pi)$. Both strategies are implemented into
a computer program.

\noindent{\bf Computer Simulation:} The master equation (\ref{MEQ}) is
solved by using minimal process algorithm suggested by
Gillespie~\cite{Gillespie1,Gillespie2}. Given that the system is in a
certain configuration, one can calculate distribution of waiting time
for the next process to happen. Time is updated by amount of waiting
$\Delta t$. Process is chosen randomly according to weight given by
corresponding rates for each process. For problem at hand a linear
selection algorithm is used.

Given that at the time $t$ the system was in the conformation ${\bm
c}$ specified in (\ref{vecc}), following processes can
happen. Transport of particle $X_\alpha$ from $C_i$ to $C_j$ occurs with
rates
\begin{equation}
  R^\alpha_{i,j}({\bm c}) = D^\alpha_{i,j} n_{\alpha,i} \ , \ \ 
  \alpha=1,2,...,M \ , \ \ i,j=1,...,N
\end{equation}
or reactions within containers with rates
\begin{equation}
  R^i_{\alpha,\beta}({\bm c})=\lambda^i_{\alpha,\beta}({\bm n}_i) n_{\alpha,i}
\end{equation}
One also needs the total reaction rate,
\begin{equation}
  Q({\bm c}) = 
     \sum_{i,j=1}^N \sum_{\alpha=1}^M R^\alpha_{i,j}({\bm c}) +
     \sum_{i=1}^N \sum_{\alpha,\beta=1}^M R^i_{\alpha,\beta}({\bm c}) 
\end{equation}
which is used to calculate the time update $\Delta t=-\ln(1-r)/Q$
where $r$ is a random number drawn uniformly from the interval
$[0,1)$. The probabilities for process to happen are given by
\begin{eqnarray}
  & & p^\alpha_{i,j}=\frac{R^\alpha_{i,j}({\bm c})}{Q} , \  
      \alpha=1,...,M , \  i,j=1,....,N \\
  & & p^i_{\alpha,\beta}=\frac{R^i_{\alpha,\beta}({\bm c})}{Q} , \ 
      i=1,...,N , \ \alpha,\beta=1,....,M 
\end{eqnarray}
A process is chosen using the linear selection algorithm. First, a new
random number $r'$ is drawn. After that, the cumulant probabilities
are calculated through loop over all processes. The loop is stopped
when the cumulant probability exceeds $r'$ and the last process for
which this happened is executed.

\noindent{\bf Moment method:} This method relies on the matrix
representation of the master equation (\ref{MEQ}) and can not be used
to treat cases with large number of containers and particle
types. However, the method is exact and should be used when there are
enough computation resources. The method is developed in the following
section.

\section{The calculation of the average and standard deviation}
\label{sec:moment_method}

It will be shown how to calculate moments of the individual components
of ${\bm \tau}$, $\gamma_k^{(p)} \equiv \langle(\tau_k)^p\rangle$,
$k=1,\ldots,\eta$ and $p=0,1,2,\ldots,\infty$.  The more complicated
moments, e.g. $\gamma_{k,l}^{(p,q)} \equiv
\langle(\tau_k)^p(\tau_l)^q\rangle$ with $k,l=1,\ldots,\eta$ and
$p,q=0,1,2,\ldots,\infty$, could be also evaluated using technique
presented in this section, but such moments will not be considered.
The focus in on the moments that are obtained through
\begin{equation}
  \gamma_k^{(p)}({\bm\iota},{\bm\pi}) = 
     \int_0^\infty t^p\  \Gamma_k(t;{\bm\iota},{\bm\pi})\ dt
  \label{moment0}
\end{equation}
where $\Gamma_k(t;{\bm\iota},{\bm\pi})$ denotes integrated
distribution function for task $\pi_k$ given by 
\begin{equation}
  \Gamma_k(t;{\bm\iota},{\bm\pi}) \equiv 
                              \int_0^\infty 
                              \Phi(t_1,\ldots,t_n;{\bm \iota},{\bm \pi}) 
                              \prod_{m=1,\eta}^{m\ne k} dt_m
  \label{gamma1}
\end{equation}
Please note that accomplishment of each individual task is influenced
by presence of others since the particles can vanish upon
accomplishment of various tasks. This is the reason why integrated
distribution function $\Gamma_k(t;{\bm\iota},{\bm\pi})$ contains both
index of the task that statistics is sought for ($k$) and full list of
tasks being monitored (${\bm\pi}$).

\begin{figure}       
%%%%%%%%%%%%%%%%%%%%%%%%%%%%%%%%%%%%%%%%       
%\epsfxsize=8cm       
%\epsfbox{XXX.eps}
\includegraphics[width=8cm]{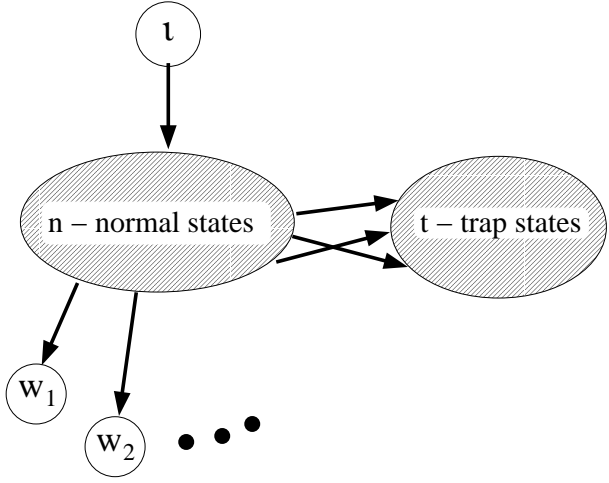}
%%%%%%%%%%%%%%%%%%%%%%%%%%%%%%%%%%%%%%%%       
\caption{The structure of the configuration space. Initially the
system is in the state ${\bm\iota}$. Three set of states are
distinguished, a set of normal states $S_n$, a set of trap states
$S_t$, and a set of window states $S_w$. See text for discussion.}
\label{diagram2}       
\end{figure}

It is possible to find closed expression for Laplace transform of
$\Gamma_k(t;{\bm\iota},{\bm\pi})$. The Laplace transform of arbitrary
function $F(t)$ is defined as $F(s)\equiv \int_0^\infty {\rm
exp}(-st)F(t)dt$. $\Gamma_k(s;{\bm\iota},{\bm\pi})$ is given by
%
%\begin{widetext}
%\begin{equation}
\begin{eqnarray}
  & & \Gamma_k(s;{\bm \iota},{\bm \pi}) = 
     \sum_{{\bm c}\ne{\bm\iota}} w({\bm c},\pi_k) 
                                 g(s;{\bm\iota},{\bm c},{\bm\pi}) \nonumber \\ 
  & &    + \sum_{m=1,\eta}^{m\ne k} \sum_{{\bm c}\ne{\bm \iota}} 
           w({\bm c},\pi_m) 
           g(s;{\bm\iota},{\bm c},{\bm\pi}) \nonumber \\ 
  & & \;\;\;\;\;\;\;\;\;\;\;\;\;\;\;\;\;\;         
          \times \Gamma_k(s;{\bm c}/\pi_m,{\bm\pi}/\pi_m) 
  \label{gamma2}
\end{eqnarray}
%\end{equation}
%\end{widetext}
%
where notation used is as follows. The $w({\bm c},\pi_k)$ equals one
if task $\pi_k$ can be accomplished once system arrives in the state
${\bm c}$, and equals zero otherwise. In the following, by definition,
a state for which one of the $w({\bm c},\pi_k)$ with $k=1,\ldots,\eta$
differs from zero will be referred to as a {\em window} state. Through
window states tasks can be accomplished. ${\bm c}/\pi_m$ denotes the
state immediately after the particles have been taken away once the
task $\pi_m$ was accomplished. Likewise, the symbol ${\bm\pi}/\pi_m$
denotes a list of tasks being monitored with task $\pi_m$ omitted
$(\pi_1,\ldots,\pi_{m-1},\pi_{m+1},\ldots,\pi_{\eta})$.
$g(s;{\bm\iota},{\bm c},{\bm\pi})$ is a distribution function for the
first passage time into the state ${\bm c}$ given that
the dynamics started from the state ${\bm\iota}$. Fig.~\ref{diagram2} is a
schematic presentation of the Eq.~(\ref{gamma2}). Please note that
Eq.~(\ref{gamma2}) defines $\Gamma_k(s;{\bm \iota},{\bm \pi})$
recursively. The number of tasks on the right hand side of
Eq.~(\ref{gamma2}) is by one smaller than the same number on the left
hand side of the equation. When list of a tasks is empty
${\bm\pi}={\bm\pi}_0$ and \mbox{${\bm\pi}_0\equiv(\ )$}. Condition
$\Gamma_k(s;{\bm \iota},{\bm\pi}_0)=0$ stops recursion.

The Laplace transform of the first arrival time distribution function
$g(t;{\bm\iota},{\bm c},{\bm\pi})$ is calculated as follows.  Given
the particular reaction scheme it is possible to construct master
equation (\ref{MEQ}) that governs the time dependence of the
occupation probabilities of each state $p({\bm c},t)$, where ${\bm c}$
has to be accessible from the initial state ${\bm\iota}$.  When calculating
matrix of the transition rates ${\bm R}$ it is assumed that all
window-states can not be left once they are arrived into. Window
states are perfectly absorbing. Fig.~\ref{diagram2} is a graphic
representation of this fact.  Once $p({\bm c},t)$ is found from
(\ref{MEQ}) the first passage time distribution function is given by
$g(t,{\bm\iota},{\bm c},{\bm\pi})=\dot p({\bm c},t)$.

It is useful to arrange both $p({\bm c},t)$ and $g(s;{\bm\iota},{\bm
c},{\bm\pi})$ into a vectors ${\bm p}(t)$ and ${\bm g}(t)$ where
notation was simplified a bit since we assume that ${\bm\iota}$ and
${\bm\pi}$ are known and fixed.  It is useful to rewrite master
equation (\ref{MEQ}) in a matrix form as $\dot{\bm p}(t) = {\bm R p}(t)$.
This equation is solved using Laplace transform, with initial
condition $p({\bm c},0)=\delta_{{\bm c},{\bm\iota}}$ 
($\delta$ denotes Kronecker delta symbol): $s {\bm p}(s)-{\bm p}_0
= {\bm Rp}(s)$. Also, in the Laplace transform space one has $s {\bm
p}(s)-{\bm p}_0 = {\bm g}(s)$, which directly leads to the equation for
the first arrival time distribution function: 
\begin{equation}
  s{\bm g}(s) = {\bm R}[{\bm g}(s)+{\bm p}_0]
  \label{gs}
\end{equation}
In principle, the equation above could be solved as ${\bm
g}(s)=(s-{\bm R})^{-1}{\bm p}_0$. However, matrix ${\bm R}$ has zero
eigenvalues and the value of ${\bm g}(s)$ in the limit $s\rightarrow
0$ is ill defined. To avoid such problems Eq.~(\ref{gs}) has to
be solved in a special way.

It is useful to separate configuration space into three groups, as shown in
Fig.~\ref{diagram2}. First group, labeled $S_n$, contains states that
are non-window or the normal-states. Second group contains states labeled
by $S_t$ that we refer to as the trap states. The third group consists of
the window states solely, labeled by $S_w$. The existence of the trap states
is problem dependent. Once system arrives into these states there is
no exit from this space, though such states are not window
states. This simply means that it is possible that set of tasks is
never accomplished.

Using the partition of states shown in Fig.~\ref{diagram2} leads to
the following set of equations
\begin{eqnarray}
   & & s {\bm g}_n(s) = {\bm R}_{nn} [{\bm g}_n(s)+{\bm p}_{n,0}]  
                        \label{gs1} \\
   & & s {\bm g}_t(s) = {\bm R}_{tn} [{\bm g}_n(s)+{\bm p}_{n,0}] 
                        + {\bm R}_{tt} {\bm g}_t(s)  
                        \label{gs2} \\
   & & s {\bm g}_w(s) = {\bm R}_{wn} [{\bm g}_n(s)+{\bm p}_{n,0}]       
                        \label{gs3}
\end{eqnarray}
Please note that ${\bm p}_{t,0}$ and ${\bm p}_{w,0}$ are zero since,
initially, the system is in the state ${\bm\iota}$ and such state does
not have any components in the $S_t$ and $S_w$ spaces. Given that
there are no transition from trap states into normal states or window
states blocks ${\bm R}_{nt}$ and ${\bm R}_{wt}$ are missing in the
equations above. Likewise, blocks ${\bm R}_{ww}$, ${\bm R}_{nw}$ and
${\bm R}_{tw}$ and are zero since there are no transitions among
window states, nor transitions from them.

The solution of the equations (\ref{gs1})-(\ref{gs3}) can be found in a
straight forward manner. Equation (\ref{gs1}) can be solved first,
leading to ${\bm g}_n(s)=(s-{\bm R}_{nn})^{-1}{\bm p}_{n,0}$ and
inserting this expression into the Eq.~(\ref{gs3}) gives
\begin{equation}
  g(s,{\bm \iota},{\bm c},{\bm \pi})=
      \left[ 
        {\bm R}_{wn} (s-{\bm R}_{nn})^{-1} {\bm p}_{n,0} 
      \right]_{\bm c} \ , \ \ {\bm c}\in S_w
  \label{gw}
\end{equation}

Please note that Eq.~(\ref{gw}) is well defined for all values of
s. In particular, in the limit $s\rightarrow 0$ even for a matrix ${\bm
R}$ that has zero eigenvalues.  It is intuitively clear that, contrary
to ${\bm R}$, matrix ${\bm R}_{nn}$ does not have zero eigenvalues:
as time goes on, all probability accumulates in $S_w$ and $S_t$ spaces
(see Fig.~\ref{diagram2}). The only difficulty with Eq.~(\ref{gw}) is
partitioning of the full configuration space into $S_n$, $S_w$ and
$S_t$. The algorithm for carrying out such partitioning is not
presented here in order to save the space.

Finally, once $g(s,{\bm \iota},{\bm c},{\bm \pi})$ is found one can
proceed with the calculation of the moments
$\gamma_k^{p}({\bm\iota},{\bm\pi})$. These can be obtained by taking
derivatives of Eq.~(\ref{gamma2}) with regard to $s$ and setting $s=0$
at the end: it can be seen easily from Eq.~(\ref{moment0}) that
\begin{equation}
  \gamma_k^{p}({\bm\iota},{\bm\pi})=
     (-)^p\lim_{s\rightarrow 0}\partial_s^p\Gamma_k(s,{\bm\iota},{\bm\pi})
  \label{gamma3}
\end{equation}
where $\partial_s$ denotes derivative over $s$. Using
Eqs.~(\ref{gamma3}) and (\ref{gamma2}) leads to
%
%\begin{widetext}
%\begin{equation}
\begin{eqnarray} 
 & & \gamma_k^{(p)}({\bm \iota},{\bm \pi}) = 
     \sum_{{\bm c}\ne{\bm\iota}} w({\bm c},\pi_k) 
        g^{(p)}({\bm\iota},{\bm c},{\bm\pi}) \nonumber \\ 
 & &  + \sum_{m=1,\eta}^{m\ne k} \sum_{{\bm c}\ne{\bm \iota}} 
           w({\bm c},\pi_m) 
           \sum_{q=0,p} \left( \begin{array}{c}
                                  p \\ q
                               \end{array} \right)
                         g^{(p)}({\bm\iota},{\bm c},{\bm\pi}) \nonumber \\
 & &  \;\;\;\;\;\;\;\;\;\;\;\;\;\;\;\;\;\;\;
      \times \gamma_k^{(p-q)}({\bm c}/\pi_m,{\bm\pi}/\pi_m) 
   \label{gamma4}
\end{eqnarray}  
%\end{equation}
%\end{widetext}
%
where by definition $g^{(p)}({\bm\iota},{\bm c},{\bm\pi})\equiv
(-)^p\lim_{s\rightarrow 0}\partial_s^p g(s,{\bm\iota},{\bm
c},{\bm\pi})$, which after using Eq.~(\ref{gw}) leads to
\begin{equation}
 g^{(p)}({\bm\iota},{\bm c},{\bm\pi}) = (-)^{(p+1)}p! 
        \left[ 
          {\bm R}_{wn} {\bm R}_{nn}^{-(p+1)} {\bm p}_{n,0} 
        \right]_{\bm c} 
 \label{gder}
\end{equation}

Equations~(\ref{gamma4})-(\ref{gder}) are central result of this
section. They determine all moments. For example, once
$\gamma_k^{(p)}({\bm\iota},{\bm\pi})$ $p=0,1,2$ are found the average
and variance of ${\bm\Tau}$ are given by
\begin{eqnarray}
 \tau_k({\bm\iota},{\bm\pi}) & = & 
    \frac{\gamma_k^{(1)}({\bm\iota},{\bm\pi})}
         {\gamma_k^{(0)}({\bm\iota},{\bm\pi})} \\
 \sigma_k({\bm\iota},{\bm\pi})^2 & = & 
    \frac{\gamma_k^{(2)}({\bm\iota},{\bm\pi})}
         {\gamma_k^{(0)}({\bm\iota},{\bm\pi})} -
    \left[\frac{\gamma_k^{(1)}({\bm\iota},{\bm\pi})}
               {\gamma_k^{(0)}({\bm\iota},{\bm\pi})}\right]^2 
\end{eqnarray}
where $k=1,\ldots,\eta$.  One has to divide by
$\gamma_k^{(0)}({\bm\iota},{\bm\pi})$ in the equations above in order
to ensure that in the case when there is a possibility that some tasks
are not accomplished statistics is done only for instances where task
was achieved. The percentage of cases when this happened is given by
$\gamma_k^{(0)}({\bm\iota},{\bm\pi})$. The numerical implementation of
Eqs.~(\ref{gamma4}) and (\ref{gder}) is a straight forward and gives
the {\em exact} values for ${\bm\tau}$ and~$\bm\sigma$.

\section{ROGE ensemble: Describing organism performance in 
terms of the single variable $\nu$}
\label{sec:ROGE1}

The calculation of ${\bm\tau}$ and ${\bm\sigma}$ for an arbitrary
organism was discussed in the previous section. In this section the
GORE ensemble will be studied. The case $N=1$ where there is only one
container is not interesting since such space is not structured.  The
first non-trivial example of a structured space is the case of a
two-container network with $N=2$. For simplicity reasons, a situation
will be considered where there are only two particle types A and B
corresponding to $M=2$. With only one particle type one can only focus
on transport issues. To see coupling between reactions and transport
one needs at least two particles types. Please note that the analysis done
in this section is quite generic, though it is carried out on a rather
simple case of $N=2$ and $M=2$. The analysis could be easily repeated for
an arbitrary values of $N$ and $M$. However, the computational cost
scales with $N$ and $M$, and there is clearly an upper limit to which
cases one can study. It will be shown that the relatively simple
two-container network and reactions with two particle types provide
interesting insight into the problem.

The structure of the organisms in the ROGE ensemble is defined as follows.
For each organism in the ensemble an unique choice is made for (i) total
number of A and B molecules, (ii) reactivity matrix $\bm\Lambda$,
(iii) catalytic activity matrix $\bm K$, (iv) inject pattern
$\bm\iota$ and (v) list of tasks monitored $\bm\pi$. For all organisms
geometry is kept fixed (e.g. size of containers and length of the tube
connecting them). Please note that there are two symmetries in the
problem, the one that originates from relabeling of particles, and
another one that has to do with relabeling of containers. A special
care is taken to eliminate these symmetries in the ensemble. The goal
is to unearth best reaction scheme (organism) that draws most benefit
from the structured space that has the form of a two-container
network. 

We start with the situation where the total number of particles
$N_p^*=n_{A,1}+n_{A,2}+n_{B,1}+n_{B,2}$ in the system equals
one. Also, we consider only organisms with reactions $A\rightarrow B$
and $B\rightarrow A$, both with rate $\lambda$. One can see that, for
a given reaction scheme, and with a constraint $N_p^*=1$, only three
choices for $\bm\iota$ and $\bm\pi$ are possible, leading to the three
organisms $o_1$, $o_2$ and $o_3$ that are listed in table
\ref{tab:simple_examples}. Please note that all other $N_p^*=1$ cases
can be related to these three through relabeling of particles and
containers, or by considering different reaction schemes.

\begin{figure}       
\psfrag{bartau}{$\tau$}
\psfrag{sigma}{$\sigma$}
\psfrag{invD}{$D^{-1}$}
%%%%%%%%%%%%%%%%%%%%%%%%%%%%%%%%%%%%%%%%       
%\epsfxsize=8cm       
%\epsfbox{histogram.eps}
\includegraphics[width=8cm]{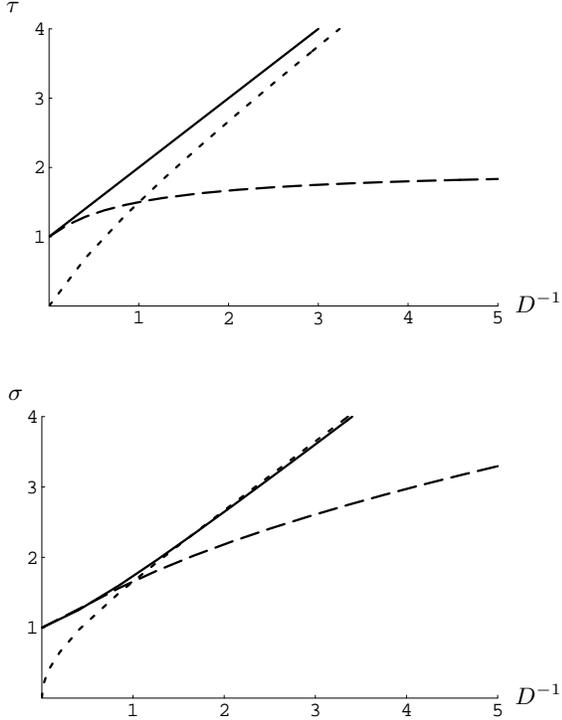}
%%%%%%%%%%%%%%%%%%%%%%%%%%%%%%%%%%%%%%%%       
\caption{The dependence of $\tau$ and $\sigma$ on $D^{-1}\sim\sqrt{l}$
for three organisms defined in table \ref{tab:simple_examples} ($l$ is
the length of a tube connecting containers).  Curves describe $o_1$
(dotted line), $o_2$ (dashed) and $o_3$ (solid).  See table
\ref{tab:simple_examples} and section \ref{sec:ROGE1} for more
details.}
\label{simple_examples}       
\end{figure}

All organisms are such that one A particle is injected in the
container $C_1$ and only one task is monitored. In the case of
organism $o_1$ the goal is to synthesize one A molecule in the
container $C_2$, in the case of $o_2$ one B molecule should be
synthesized in $C_1$, while in the case of $o_3$ one B molecule should
be created in $C_2$.  Reaction process is stochastic. Average and
standard deviation of time for task accomplishment are given under
columns labeled $\tau$ and $\sigma$ in table
\ref{tab:simple_examples}. The dependence of $\tau$ and $\sigma$ on
(the inverse of) the transport rate $D$ is depicted in
Fig.~\ref{simple_examples}.

Clearly, all three organisms achieve their tasks faster in the compact
geometries. This can be seen from Fig.~\ref{simple_examples} since
$\tau$ gets smaller when $D^{-1},l\rightarrow 0$. (Here, we used the
fact that $D^{-1}\sim l^2$ where $l$ is the length of the tube
connecting containers). Same holds for curves depicting
$\sigma$. Amount of noise decreases when geometry is compact. For
example, organism $o_1$ functions by transporting one A particle from
container $C_1$ to container $C_2$. It is intuitively clear that this
happens faster when containers are close. One can analyze $o_3$ in the
similar way. However, $o_2$ is somewhat different. In the case of
$o_2$ the goal is to synthesize one B molecule in the same container
where A was injected. Figure \ref{simple_examples} illustrates the
fact that when another open volume is present molecule can wonder away
into additional volume and this process delays synthesis of B. One can
also see that $o_2$ is least sensitive to increase (decrease) in $l$
($D$): when $l,D^{-1}\rightarrow\infty$ the $\tau$ for $o_1$ and $o_3$
increases, while for $o_2$ $\tau$ saturates to constant (though level
of noise, $\sigma$, increases).

The question is whether it is possible to present information conveyed
from Fig.~\ref{simple_examples} and table~\ref{tab:simple_examples} in
a more compact way? There are couple of reasons for this. First, in a
case of the more complicated geometry, generating graph such as shown
in Fig.~\ref{simple_examples} is time consuming. Second, it would be
desirable to have automatic procedure for assessing most important
properties of such graph; whether given organism performs best in
compact or structured (network like) geometry.

To achieve this goal and compactify information in
Fig.~\ref{simple_examples} and table~\ref{tab:simple_examples} it is
useful to consider following quantity,
\begin{equation}
  \nu = \frac{\|{\bm\tau}_n\|}{\|{\bm\tau}_0\|}
\end{equation}
where $\|.\|$ denotes Euclidean norm of the vector, ${\bm\tau}_n$ and
${\bm\tau}_0$ are given by ${\bm\tau}$ calculated for network like and
compact geometries with jump rates $D_n$ and $D_0$ such that
$D_n\sim\lambda$ and $D_0\gg D_n,\lambda,\xi\lambda$.  Using similar
reasoning the $\nu$ can be defined for generic network having more
than two containers. In such a way one can compare extended and
compact geometries of given fixed network structure and express
comparison through one variable~$\nu$.

In the following the $\nu$ will be refereed to as {\em speed} of
a reaction. $\nu<1$ is indication that organisms accomplishes tasks
faster in a network like geometry, while $\nu>1$ shows that organism
draws most benefit from a compact geometry. Please note, all organisms
considered in table \ref{tab:simple_examples} and
Fig.~\ref{simple_examples} have $\nu>1$. 

In principle, one could define quantity similar to $\nu$
and use $\sigma$ instead of $\tau$. For example, one could use
\begin{equation}
  \mu = \frac{\|{\bm\sigma}_n\|}{\|{\bm\sigma}_0\|}
\end{equation}
where meanings of ${\bm\sigma}_n$ and ${\bm\sigma}_0$ are similar to
the ones of ${\bm\tau}_n$ and ${\bm\tau}_0$. $\mu$ could be used to
classify organisms in the ensemble according to the amount of noise in
compact and extended (network-like) geometries. Also, it is possible
to use $\rho=\sqrt{\nu^2+\mu^2}$ as simultaneous measure of speed and
noise. But such quantities will not be studied at the moment. From now
on we focus on $\nu$ solely.

\section{ROGE ensemble: Classification of the reaction 
schemes using the speed of reaction $\nu$} 
\label{sec:ROGE2}

Figure \ref{histogram} depicts results of the classification of large
number of organisms in five ROGE ensembles where $\nu$ (defined in
previous section) is used as a measure of the performance. Five ROGE
ensembles are constructed with the increasing upper limit for the
total number of particles in the system $N_p^*$ from $1$ to $5$. 

For $N_p^*=1$ there are 95 unique organisms. Clearly, this number is
overestimate. The number of unique organisms is obtained after
eliminating symmetries related to relabeling of particles and
containers removed. However, this number is still too large since
catalytic influences are assumed to play the role, though there can
not be any catalytic influence when there is constantly one molecule
in the system.  The number of unique organisms for other cases
discussed later with $N_p^*=2,3,4,5$ is correct.  There are no
organisms in the $N_p^*=1$ class that benefit from the structured
geometry since the histogram in the panel (b) is empty. All organisms
have $\nu>1$ as can be seen from panel (a).

For $N_p^*=2$ there are 730 organisms that are unique. Only after more
than one molecule appears in the system, catalytic influences start to
play the role, and organisms that benefit from the network like structure
appear [see panels (c) and (d) in Fig.~\ref{histogram}]. The best
performer in this class is given in Table~\ref{tab:best}, denoted by
$o_{i,2}$, together with couple of a second best performers denoted by
$o_{ii,2}$. The two organisms labeled $o_{i,2}$ draw most benefit from
the network like structure and degeneracy in $\nu$ comes from the fact
that same enhancement and reduction factor is used for positive and
negative catalytic influence respectively.

One can understand intuitively why $o_{i,2}$ runs faster in the
network like geometry. We focus on the particular case of $o_{i,2}$
where goal is to synthesize two B molecules in the container
$C_1$. Due to the initial presence of molecule A in the container
$C_1$, it is very likely that B molecule will be converted into A
molecule, when A and B meet in the same container. Once there are two
A molecules in the system the trouble starts. Even if the reaction
$A\rightarrow B$ happens, and chance for this is really small due to
the negative catalytic influence of A on such reaction, B will be
converted back to A immediately (due to the positive catalytic
influence of another A molecule on the $B\rightarrow A$ reaction).  In
principle, conversion of AB into 2B has no chance occurring in a
reasonable time when there is only one container. The antagonistic
catalytic influences just discussed are best handled in a network like
geometry. In such a case the synthesis of the molecules can be done in
a separate containers. With two containers there is always a chance
that antagonistic influences will be reduced. For example, in the case
there are two A molecules in the system, the processes $A\rightarrow
B$ can happen fast given that damage inflicting A molecule is in
another container.

Also, it is naive to think that organisms containing reactions with
solely negative catalytic influence perform best in a network like
structure. This is clearly not the case.  The winning organisms
$o_{i,2}$ are constructed from one reaction with positive
$B\stackrel{+A}{\longrightarrow}A$ and another reaction 
$A\stackrel{-A}{\longrightarrow}B$ with negative catalytic influence.
Furthermore, Fig.~\ref{histogram} shows plenty of other cases.
Actually, the completely opposite is possible. There are many
organisms in the histogram plot with $\nu>1$ that contain at least one
reaction with negative catalytic influence and these organisms perform
best in a compact geometry.

With $N_p^*=3$ there are 3025 organisms that are unique.  The best
performer in this class is given in Table~\ref{tab:best}, denoted by
$o_{i,3}$. A couple of the second best performers are also shown and
labeled $o_{ii,3}$. When more than two molecules appear in the system,
completely new organism appears as winner. There is a sharp transition
in character of winners. Both the reaction type and inject and task
patterns are different in $o_{i,3}$ and $o_{i,2}$. The reaction
of organisms $o_{3,i}$ is such that it is possible that tasks are not
accomplished. For example, all A molecules can be converted to B
before task [2A]-\cz\ in $C_1$ is accomplished. Once this happens
synthesis of A's in $C_1$ will never occur since back reaction
$B\rightarrow A$ is absent. 

Very likely, from biological point of view there is no advantage
basing survival on a reaction that sometimes fails, i.e. using
reaction similar to the one contained in $o_{i,3}$. Nevertheless,
there might be constraints that could enforce presence of such
reaction. For example, the number of molecules in Nature is large but
limited. Living organisms have to use what is available. In the lack
of alternatives it might be necessary for the intra cellular machinery
to use reaction that sometimes fails. But leaving such discussions
aside, the only point emphasized here is that organism $o_{i,3}$ draws
most benefit from the network like structure.

Interestingly, this type of organism remains winner in the classes
$N_p^*=4$ and $N_p^*=5$ with $\nu=0.0248$ and $\nu=0.0192$, see
table~\ref{tab:best}, organisms $o_{i,4}$ and $o_{i,5}$. However,
please note that details of the inject pattern of $o_{i,4}$ and
$o_{i,5}$ are somewhat different from $o_{i,3}$. Judging solely from
$o_{i,3}$ and $o_{i,4}$ one would guess that inject patter for best
performer in class $N_p^*=5$ should be ${\bm\iota}=$[A]$-$[4A], in
class $N_p^*=6$ ${\bm\iota}=$[A]$-$[5A], etc. However, this is not the
case. The best performer in class $N_p^*=5$ has inject pattern equal to
${\bm\iota}=$[2A]$-$[3A] with $\nu=0.0192$, while organism with
${\bm\iota}=$[A]$-$[4A], denoted by $o_{ii,2}$, has somewhat larger
$\nu=0.0219$.

Thus, examples above show how difficult it is to have any intuition
about structure of best performers. There are also other ways to see
this. For example, the last row in table~\ref{tab:best} contains
organisms with slightly modified inject or task patterns where
original form is taken from best performer. Comparing $o_{i,3}$ and
$o_{*,3}$, $o_{i,4}$ and $o_{*,4}$, and finally $o_{i,5}$ with
$o_{*,5}$ shows that small alternation in task pattern, obtained by
moving one B particles from $C_2$ into $C_1$, lowers performance
considerably. Organisms with such alternations do not perform well in
the network like geometry: each of $o_{*,3}$, $o_{*,4}$ and $o_{*,5}$
has $\nu>1$. Another example, can be obtained from comparison of
$o_{i,3}$ with $o_{*,2}$. Both inject and task patterns have been
altered in $o_{i,3}$. This change is motivated by sequence of
organisms in first row of table~\ref{tab:best}, when read from right
to left. A priori, the organism $o_{*,2}$ could be considered to have
$\nu<1$, however this is not the case. The actual value for $\nu=418$
is (lot) larger than one.

In summary, the table~\ref{tab:best} shows a couple of interesting
features. First, one can see that when maximum allowed number of
particles in the system increases new effects appear. It is impossible
to predict winner in each class before calculation is done. Second,
there is also a large sensitivity on inject and task patterns. Slight
alternation of these patterns can lead to drastic changes in
performance criteria $\nu$.

\section{Discussion}
\label{sec:discussion}

We introduced what we might call a generic model for study of chemical
reactions in structured spaces, based on a simple way of incorporating
interplay between transport, chemical reactions, and geometry. A
number of different chemical reactions were studied and their
performance compared in various ways.  The main idea is to see how the
reaction processes behave when geometry changes from compact open
space towards the cell like environment that is more structured. In
order to be able to analyze large number of reactions, relatively
simple model was used in order to reduce computational
time. Following simplifications were made.

As a model of structured space we used a network consisting of
containers connected by tubes.  Such setup captures the most important
geometrical aspects of the problem. There are compact regions in space
that reactants can explore and react within. These regions are
connected to each other and exchange chemicals. In such a way the 
two most important processes are clearly separated, the local dynamics
within container, and the transport among containers.  This is pretty much
how living cell operates. In such a way one can capture most important
geometrical aspects of the cell interior.

The reaction scheme considered is rather crude.  Dynamics
within container, however complicated it may be, is projected onto a
relatively low number of degrees of freedom, the number of particles
in each container. The reaction rate $\lambda$ describes how fast
reactions happen. The description in terms of number of particles
becomes more and more valid when size of the reaction volume is
reduced and number of reactants few~\cite{McQuarrie1,Clifford1}.  It
is exactly this limit that we focussed on.

The transport between containers was modeled in simplest possible way
by using the effective transport rate $D$. It was assumed that
transport rate is same for any pair of containers, and type of
particles. This assumption could be easily relaxed. Clearly, when
tubes are long, the number of particles in each container decays
non-exponentially and the transport process can not be described by
using a transport rate. But such effects are not considered here.

Thus, only two parameters are used to describe dynamics, the reaction
rate $\lambda$ and transport rate $D$. There are several advantages in
doing so, but the most important one is that reactions and transport
are clearly separated, and it is easier to understand how they
influence each other.  Despite apparent simplicity, the model studied
in here captures all essential features of the problem. Should there
be need for that one can make the model more realistic, but we refrain
from this at the moment.

Which types of chemical reactions draw most benefit from structured
spaces?  In an attempt to answer this question two schemes were
formulated. In ROGE scheme one explores variety of chemical reactions
while spatial structure (geometry) is kept fixed. It is the other way around
for GORE scheme. At the moment ROGE scheme is studied in a lot more
detail. The study of GORE ensemble will be left for the future work. The
main findings of this work are discussed bellow.

Since dynamics is stochastic one needs probabilistic description of
the system, and in order to perform analysis of any chemical reaction
one has to solve corresponding master equation that describes
statistics of events.  Two techniques for doing this were used. First,
the computer simulation is a straight forward way to analyze master
equation (discussed in section~\ref{sec:dynamics}). This technique 
is not that accurate due to the stochastic spread of data. One needs
unrealistically large number of runs in order to gain reasonable
accuracy in $\nu$. In here, a novel method of analyzing chemical
reaction was developed, with emphasis on the first passage time (see
section~\ref{sec:moment_method}). This method relies on a matrix
representation of the master equation. In principle, it is exact up to
the numerical errors in carrying out matrix manipulations, such as
finding inverses, multiplying matrix with vector etc. In here we
developed a equation that describes first passage time distribution
function for achieving set of tasks. From this equation one can easily
obtain all moments of distribution function and calculate average
execution time for reaction and its variance.

It is not so easy to compare two vastly different chemical reactions.
A method for doing this was discussed in
section~\ref{sec:ROGE1}. There are some obvious difficulties when
doing comparison. For example, one might need to compare a situation
where number of molecular types involved in reactions are completely
different. Also, spatial structures need not be the same, e.g. there
might be a need to compare two vastly different network structures. In
order to overcome these difficulties one has to use a measure of
relative performance. For example, in the case of a ROGE ensemble the
execution time for given reaction was compared when geometry is
stretched (transport occurs with rate $D_n\sim\lambda$) and compact
(transport with rate $D_0$ where $D_0\gg D_n,\lambda,\xi\lambda$). The
ratio $\nu$ of the magnitude of the execution times ${\bm\tau}_n$ and
${\bm\tau}_0$ for stretched (network) and compact geometry is a rough
estimate of how well reaction performs in a network like
geometry. Instead of comparing organisms directly one compares values
$\nu=\|{\bm\tau}_n\|/\|{\bm\tau}_0\|$.  

\begin{figure}       
\psfrag{noo}[][]{\small{\#o}}
\psfrag{ npmax1}{\small{$N_p^*=1$}}
\psfrag{ npmax2}{\small{$N_p^*=2$}}
\psfrag{ npmax3}{\small{$N_p^*=3$}}
\psfrag{ npmax4}{\small{$N_p^*=4$}}
\psfrag{ npmax5}{\small{$N_p^*=5$}}
\psfrag{t.t0}{\small{$\frac{\tau_n}{\tau_0}$}}
\psfrag{t0.t}[][]{\small{$\chi\left(\frac{\tau_0}{\tau_n}\right)$}}
\psfrag{A}{(a)}
\psfrag{B}{(b)}
\psfrag{C}{(c)}
\psfrag{D}{(d)}
\psfrag{E}{(e)}
\psfrag{F}{(f)}
\psfrag{G}{(g)}
\psfrag{H}{(h)}
\psfrag{I}{(i)}
\psfrag{J}{(j)}
%%%%%%%%%%%%%%%%%%%%%%%%%%%%%%%%%%%%%%%%       
%\epsfxsize=8cm       
%\epsfbox{histogram.eps}
\includegraphics[width=8cm]{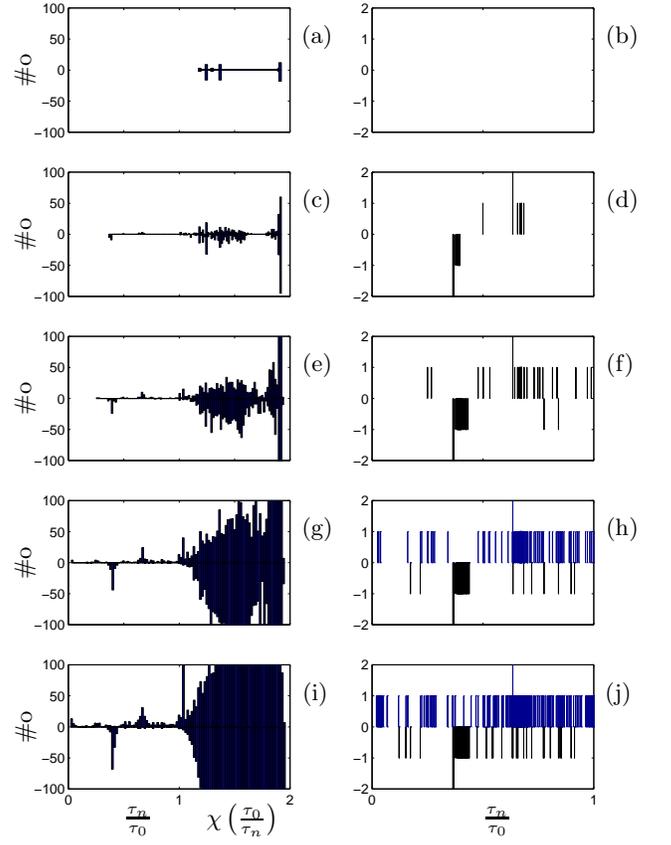}
%%%%%%%%%%%%%%%%%%%%%%%%%%%%%%%%%%%%%%%%       
\caption{Classification of organisms in ROGE ensemble. $\nu$ was
calculated with $D_n=1s^{-1}$, $D_0=3125s^{-1}$, $\lambda=1s^{-1}$,
and $\xi=100$. Panels (a), (c), (e), (g) and (i) are histograms that
depicts groups with similar reaction speed $\nu$. There are 100
classes of width $0.01$ for $\nu$ from $1$ to $2$. For $\nu>1$,
instead of $\nu$, the value obtained from $\chi(\nu)=2-\ln
2/\ln(1+\nu)$ is used. The function $\chi$ monotonically grows with
$\nu$ and maps infinite interval $[1,\infty)$ onto the finite one
$[1,2)$. In addition, this particular form for $\chi$ reveals more
details in the region near $\nu=1$. Panels (b), (d), (f), (h), and (j)
are discrete spectra (no histogram) for region $\nu\in[0,1]$. Panels
in the same row have same value for the total number of particles in
the system $N_p^*$: (a) and (b) $N_p^*=1$, (c) and (d) $N_p^*=2$, (e)
and (f) $N_p^*=3$, (g) and (h) $N_p^*=4$, (i) and (j)
$N_p^*=5$. Negative value for number of organisms ($\#o$) indicates
that organism with particular value of $\nu$ contains at least one
reaction that can be influenced through mechanism of
suppression. Presence of suppressor lowers the reaction rate from
$\lambda$ to $\lambda/\xi$ (see section~\ref{sec:model} for details).}
\label{histogram}       
\end{figure}

Thus, the single variable $\nu$ is sufficient to determine whether a
reaction (organism) runs best on the network like ($\nu<1$) or the
compact geometry ($\nu>1$). The variable $\nu$ was used to classify
performance of various organisms in the geometry consisting of two
containers connected by tube, with main findings summarized in
Fig.~\ref{histogram} and Table~\ref{tab:best}.

(1) It is obvious that intuition does not help much. One really has
to do numerical analysis in order to extract best performer in a given
class. For example, the character of best performer changes quite a
bit when number of particles in the system varies.  The structure of
the winning organism in the first row of table changes in a rather
unpredictable manner as upper limit to the total number of particles
in the system increases from 2 to 5.  Also, the performance is
extremely sensitive to the details of inject and task patterns.  It is
interesting to speculate whether such sensitivity on total particle
number, inject and task patterns is exploited in the intra cellular
machinery. We developed a quantitative method to judge on such
effects.

(2) The role of reactions with positive and negative catalytic
influence is symmetric. Roughly, they occur equally often in the
organisms that perform well in network like geometries with
$\nu<1$. Interestingly enough, same holds for opposite range with
$\nu>1$. From Fig.~\ref{histogram} one can see that positive and
negative areas are roughly equal in magnitude, both in the $\nu<1$ and
$\nu>1$ regions. Thus, there are reactions of the suppressor type that
run better in compact geometries, and the other way around. There are
also reactions with solely positive catalytic influence that run
faster in network like geometry. But these findings are the not the
only ones that are counterintuitive. It is also surprising that in the
region $\nu\approx 0$ reactions with solely positive catalytic
influence dominate.

(3) Reaction schemes containing antagonistic catalytic influences in
the intermediate stages of reaction, that slow down the production of
the final product, require network like geometry to run fast. The
antagonistic sub-reactions have to isolated and allowed to occur in
separate regions of space. There are two mechanisms that work against
each other. First, the reaction time gets larger since one needs to
transport reactants to the regions where damage inflicting
sub-reactions must be run in isolation. Second, after being isolated,
dangerous sub-reactions happen much faster than when all reactants are
mixed and this shortens reaction time. It is interesting to note that
there are large number of cases where isolating misbehaving
sub-reactions pays off in faster reaction time.

(4) In general, antagonistic catalytic influences are hard to
identify.  It is hard to judge reaction by itself. The whole triple
consisting of reaction, inject pattern, and task pattern has to
be considered simultaneously.

In summary, we studied the workings of a chemical reactions in a two
vastly different spaces. In the compact conformation, molecules can
reach any part of the space very fast. The transport rate is lot
larger than any reaction rate in the system. In the network like
conformation, various volumes are well separated and the transport of
reactants between volumes occurs relatively slowly. It was found that
considerable number of reactions work better in a network like
configuration when transport rate gets smaller.  The setup suggested
in here is generic. There are many possible ways of extending present
analysis. For example, at present focus is on small number of
particles and stochastic dynamics. One could easily consider situation
where number of particles is larger and classical chemical kinetics
applies in the container. Transport between containers can be treated
in a better way. Instead of focusing on the average execution time
$\bm\tau$ one can easily look at noise $\bm\sigma$ or combination of
the two. The GORE ensemble should be explored in a lot more
detail. These issues will be addressed in the future work.

\begin{acknowledgments}
I would like to thank Prof. Owe Orwar and members of his group for
fruitful discussions that provided motivation for this work and for
warm hospitality during the S\"ar\"o meeting 2005. The financial
support of Prof. Owe Orwar is greatly acknowledged.
\end{acknowledgments}

%%%%%%%%%%%%%% Figure captions %%%%%%%%%%%%%%%%%%%       
         
\begin{table}
\begin{samepage}
\caption{\label{tab:simple_examples} Analysis of a three cases where
the total number of particles equals one and the number of containers
equals two. These are simplest cases that one can consider with regard
to the choices of $\bm\iota$ and $\bm\pi$. A reaction scheme used is
$A\rightarrow B$ and $B\rightarrow A$, both with rate $\lambda$
(effects of catalyst or suppressor are absent since there is only one
particle in the system at a time). The jump rate between containers is
given by $D$. In the definition of $\bm\iota$ and $\bm\pi$ the
content of containers is symbolically indicated as
\mbox{[content~of~$C_1$]-[content~of~$C_2$]}.}
\begin{ruledtabular}
\begin{tabular}{c|cc|cc}
%\hline
 org. &  $\bm\iota$ & $\bm\pi$ & $\tau$ & $\sigma^2$ \\ 
 \hline
 $o_1$ & [A]-[\ ] & [\ ]-[A] \ \  
       & $\frac{2\,D+\lambda}{D\,\left(D+\lambda\right)}$ 
       & $\frac{2\,D^3+5\,D^2\,\lambda+3\,D\,{\lambda}^2+{\lambda}^3}
               {D^2\,\lambda\,{\left(D+\lambda \right)}^2}$ \\
 $o_2$ & [A]-[\ ] & [B]-[\ ] \ \ 
       & $\frac{D + 2\,\lambda }{\lambda \,\left( D + \lambda  \right) }$ 
       & $\frac{D^3 + 3\,D^2\,\lambda  + 5\,D\,{\lambda }^2 + 2\,{\lambda }^3}
               {D\,{\lambda }^2\,{\left( D + \lambda  \right) }^2}$ \\
 $o_3$ & [A]-[\ ] & [\ ]-[B] \ \ 
       & $\frac{1}{D}+\frac{1}{\lambda}$ 
       & $\frac{1}{D^2}+\frac{1}{D\lambda}+\frac{1}{\lambda^2}$ \\
\end{tabular}
\end{ruledtabular}
\end{samepage}
\end{table}

\begin{table*} 
\caption{\label{tab:best} List of the best performers in the ROGE
ensemble (first row). Couple of second best performers are also shown
(second row). Last row contains organisms that are obtained by
slightly perturbing winners from the first row.}
\begin{ruledtabular}
\begin{tabular}{cccc}
 $N_p^*=2$ & $N_p^*=3$ & $N_p^*=4$ & $N_p^*=5$ \\
 \hline
 \begin{tabular}{c}
 ( $o_{i,2}$ ) \\
 $A\stackrel{-A}{\longrightarrow}B$, $B\stackrel{+A}{\longrightarrow}A$ 
 ($R_1$) \\  
 $\bm\iota=$[A]-[B] \\
 $\bm\pi=$[2B]-\cz\ or \cz-[2B] \\
 $\bm\tau_n=$[18.55]-\cz \\
 $\bm\tau_0=$[50.91]-\cz \\
 $\nu=0.364$
 \end{tabular} &
      \begin{tabular}{c}
      ( $o_{i,3}$ ) \\
      $A\stackrel{+A}{\longrightarrow} B$ ($R_3$)\\ 
      $\bm\iota$=[A]-[2A] \\
      $\bm\pi$=[2A]-[B] \\
      $\bm\tau_n$=[0.242]-[0.0355] \\
      $\bm\tau_0$=[0.000219]-[0.979] \\
      $\nu$=0.250 
      \end{tabular} &
           \begin{tabular}{c}
            ( $o_{i,4}$ ) \\
            $A\stackrel{+A}{\longrightarrow} B$ ($R_3$) \\ 
            $\bm\iota=$[A]-[3A] \\
            $\bm\pi=$[3A]-[B] \\
            $\bm\tau_n=$[$0.021$]-[$0.01$] \\
            $\bm\tau_0=$[$0.00045$]-[$0.95$] \\
            $\nu=0.0248$
           \end{tabular} &
                \begin{tabular}{c}
                 ( $o_{i,5}$ )  \\
                 $A\stackrel{+A}{\longrightarrow} B$ ($R_3$) \\ 
                 $\bm\iota=$[2A]-[3A] \\
                 $\bm\pi=$[4A]-[B] \\
                 $\bm\tau_n=$[0.014]-[0.00063] \\
                 $\bm\tau_0=$[0.010]-[0.91] \\
                 $\nu=0.0192$
                \end{tabular} \\
 \hline
 \begin{tabular}{c}
  ( $o_{ii,2}$ ) \\
  $A\stackrel{-B}{\longrightarrow} B$, 
  $B\stackrel{+A}{\longrightarrow} A$ ($R_2$) \\ 
  $\bm\iota=$[A]-[B],
  $\bm\pi=$[2B]-\cz\ \\ 
  or \cz-[2B],  $\nu=0.368$ \\
  $R_1$,
  $\bm\iota=$\cz-[2A], \\
  $\bm\pi=$[2B]-\cz,
  $\nu=0.386$ \\
  $R_3$,
  $\bm\iota=$\cz-[2A], \\
  $\bm\pi=$[A]-[B],
  $\nu=0.500$
  \end{tabular} &
       \begin{tabular}{c}
       ( $o_{ii,3}$ ) \\
       $R_3$, 
       $\bm\iota$=\cz-[3A] \\
       $\bm\pi$=[2A]-[B],
       $\nu=0.268$ ; \\
       $R_1$,
       $\bm\iota$=[2A]-[B] \\
       $\bm\pi$=[A]-[2B] \\
       $R_2$,
       $\bm\iota$=[2A]-[B] \\
       $\bm\pi$=[2B]-[A]\ or \cz-[A,2B] \\
       $\nu\approx 0.386$
       \end{tabular} &
           \begin{tabular}{c}
            ( $o_{ii,4}$ )  \\
            $R_3$, 
            $\bm\iota$=[2A]-[2A] \\
            $\bm\pi$=[3A]-[B], 
            $\nu$ =$0.0279$; \\
            $R_3$,  
            $\bm\iota$=\cz-[4A] \\
            $\bm\pi$=[3A]-[B],
            $\nu$=$0.0358$ 
            \end{tabular} &
                 \begin{tabular}{c}
                 ( $o_{ii,5}$ )  \\
                 $R_3$, 
                 $\bm\iota$=[2A]-[3A] \\
                 $\bm\pi$=[4A,B]-\cz, 
                 $\nu$=0.0219; \\
                 $R_3$,  
                 $\bm\iota$=[A]-[4A] \\
                 $\bm\pi$=[4A]-[B],
                 $\nu$=0.0305 
                 \end{tabular} \\
  \hline
  \begin{tabular}{c}
  ( $o_{*,2}$ ) \\
  $R_1$\ or $R_3$,
  $\bm\iota=$\cz-[2A] \\
  $\bm\pi=$[2A]-\cz,
  $\nu=$3249 or 418; \\
  %
  %$R_1$\ or $R_2$, 
  %$\bm\iota=$[2A]-\cz, \\
  %$\bm\pi=$[2B]-\cz\ or \cz-[2B] ; \\
  %
  %$R_1$\ or $R_2$, 
  %$\bm\iota=$[AB]-\cz, \\
  %$\bm\pi=$[2B]-\cz\ or \cz-[2B] ; \\
  %
  %$0.377\le\nu\le 0.396$
  \end{tabular} & 
       \begin{tabular}{c}
       ( $o_{*,3}$ ) \\
       $R_3$,
       $\bm\iota=$[A]-[2A] \\
       $\bm\pi=$[3A]-\cz,
       $\nu=21.8$
       \end{tabular} &
            \begin{tabular}{c}
            ( $o_{*,4}$ ) \\
            $R_3$,
            $\bm\iota=$[A]-[3A] \\
            $\bm\pi=$[4A]-\cz,
            $\nu=17.5$
            \end{tabular} & 
                 \begin{tabular}{c}
                 ( $o_{*,5}$ ) \\
                 $R_3$,
                 $\bm\iota=$[2A]-[3A] \\
                 $\bm\pi=$[5A]-\cz,
                 $\nu=10.4$
                 \end{tabular}  
\end{tabular}
\end{ruledtabular}
\end{table*}

%%%%%%%% References  %%%%%%%%%%%%%%%%%%%%%%%%%%       
       
%\bibliography{refs}

\begin{thebibliography}{10}

\bibitem{Nelsestuen}
G.L. Nelsestuen, Chemistry and Physics of Lipids {\bf 101}, 37 (1999).

\bibitem{Kuthan}
H. Kuthan, Progress in Biophysics \& Molecular Biology {\bf 75}, 1 (2001).

\bibitem{Ganti}
T. Ganti, BioSystems {\bf 7}, 15-21 (1975).

\bibitem{SSFA}
M.L. Simpson,G.S. Sayler, J.T. Fleming and B. Applegate, Trends in
  Biotechnology {\bf 19}(8), 317 (2001).

\bibitem{MH}
A. Mikhailov and B. Hess, J. Phys. Chem. {\bf 100}, 19059 (1996).

\bibitem{SMH1}
P. Stange, A.S. Mikhailov and B. Hess, J. Phys. Chem. B {\bf 102}, 6273 (1998).

\bibitem{SMH2}
P. Stange, A.S. Mikhailov and B. Hess, J. Phys. Chem. B {\bf 103}, 6111 (1999).

\bibitem{SZMH}
P. Stange, D. Zanette, A.S. Mikhailov and B. Hess, Biophysical Chemistry {\bf
  79}, 233 (1999).

\bibitem{SMH3}
P. Stange, A.S. Mikhailov and B. Hess, J. Phys. Chem. B {\bf 104}, 1844 (2000).

\bibitem{WQL}
H. Wang, Q. Quyang and Y. Lei, J. Phys. Chem. B {\bf 105}, 7099 (2001).

\bibitem{SQ}
K. Sun and Q. Quyang, Phys. Rev. E {\bf 64}, 026111 (2001).

\bibitem{QQ}
H. Qian and M. Qian Phys. Rev. Lett. {\bf 84}(10), 2271 (2000).

\bibitem{Calvin}
M. Calvin, {\em Chemical Evolution: Molecular evolution towards the origin of
  living systems on the earth and elsewhere}, (Claredon Press, Oxford, 1969).

\bibitem{Volkenstein}
M.V. Volkenstein, {\em Physical Approaches to Biological Evolution},
  (Springer-Verlag, 1994).

\bibitem{Langton}
C. Langton, Physica D {\bf 22}, 120-149 (1986).

\bibitem{Furusawa}
C. Furusawa and K. Kaneko, Artificial Life {\bf 4}, 79-93 (1998).

\bibitem{Vespalcova}
Z. Vespalcova, A.V. Holden, and J. Brindley, Phys. Lett. A {\bf 197}, 147-156
  (1995).

\bibitem{Ono}
N. Ono and T. Ikegami, J. Theor. Biol. {\bf 206}, 243-253 (2000).

\bibitem{Boer}
M. Boerljist and P. Hogeweg, Physica D {\bf 48}, 17 (1992).

\bibitem{Owe1}
Karlsson M, Davidson M, Karlsson R, Karlsson A, Bergenholtz J, Konkoli Z,
  Jesorka A, Lobovkina T, Hurtig J, Voinova M, Orwar O, Ann. Rev. Phys. Chem.
  {\bf 55}, 613 (2004).

\bibitem{Owe2}
Karlsson M, Sott K, Davidson M, Cans AS, Linderholm P, Chiu D, Orwar O, Proc.
  Nat. Acad Sci. USA {\bf 99}, 11573 (2002).

\bibitem{Owe3}
Karlsson A, Karlsson R, Karlsson M, Cans AS, Stromberg A, Ryttsen F, Orwar O,
  Nature {\bf 409}, 150 (2001).

\bibitem{Owe4}
Karlsson M, Nolkrantz K, Davidson MJ, Stromberg A, Ryttsen F, Akerman B, Orwar
  O, Anal. Chem. {\bf 72}, 5857 (2000).

\bibitem{Ross1}
J.P. Laplante, M. Pemberton, A. Hjelmfelt and J. Ross, J. Phys. Chem. {\bf
  99}(25), 10063 (1995).

\bibitem{Ross2}
A. Hjelmfelt and J. Ross, J. Phys. Chem. {\bf 97}, 7988 (1993).

\bibitem{Ross3}
A. Hjelmfelt, F.W. Schneider and J. Ross, Science {\bf 260}, 335 (1993).

\bibitem{Bagley1}
R.J. Bagley and J.D. Farmer, in {\em Artificial Life II}, Eds. C.G. Langton, C.
  Taylor, J.D. Farmer and S. Rasmussen, (Addison-Wesley, 1992), pp. 93-140.

\bibitem{Bagley2}
R.J. Bagley, J.D. Farmer and W. Fontana, in {\em Artificial Life II}, Eds. C.G.
  Langton, C. Taylor, J.D. Farmer and S. Rasmussen, (Addison-Wesley, 1992), pp.
  141-158.

\bibitem{Farmer}
J.D. Farmer, S.A. Kauffman and N.H. Packard, Physica D {\bf 22}, 50-67 (1986).

\bibitem{Kauffman}
S.A. Kauffman, J. Theor. Biol. {\bf 119}, 1-24 (1986).

\bibitem{Stadler}
P.F. Stadler, W. Fontana and J.H. Miller, Physica D {\bf 63}, 378-392 (1993).

\bibitem{Schuster}
P. Schuster, Physica D {\bf 22}, 100-119 (1986).

\bibitem{Slanina}
F. Slanina and M. Kotrla, Phys. Rev. Lett. {\bf 83}, 5587 (1999).

\bibitem{Jain}
S. Jain and S. Krishna, Phys. Rev. Lett. {\bf 81}, 5684 (1998).

\bibitem{Kotomin1}
E. Kotomin and V. Kuzovkov, Rep. Prog. Phys. {\bf 55}, 2079 (1992).

\bibitem{Kotomin2}
E. Kotomin and V. Kuzovkov, Comprehensive Chemical Kinetics, Vol. 34,
  R.G.Compton and G. Hancock Editors, (Elsevier, 1996), ``Modern aspects of
  diffusion-controlled reactions''.

\bibitem{CompChemKin}
Comprehensive Chemical Kinetics, Vol. 25, ``Diffusion-limited reactions'', C.H.
  Bamford, C.F.H. Tipper and R.G. Compton Editors, (Elsevier, 1985).

\bibitem{Kotomin3}
E. Kotomin and V. Kuzovkov, Comprehensive Chemical Kinetics, Vol. 34,
  R.G.Compton and G. Hancock Editors, (Elsevier, 1996), ``Modern aspects of
  diffusion-controlled reactions''.

\bibitem{Mikhailov1}
A. S. Mikhailov, Phys. Rep. {\bf 184}, 307 (1989).

\bibitem{Ovchinnikov}
A.A. Ovchinnikov, S.F. Timashev, and A.A. Belyy, ``Kinetics of diffusion
  controlled chemical processses'', (Nova Science, 1989).

\bibitem{McQuarrie1}
McQuarrie D, J. Appl. Prob. {\bf 4}, 413-478 (1967).

\bibitem{Clifford1}
Clifford P, Green NJB, Pilling MJ, J. Phys. Chem. {\bf 86}, 1318-1321 (1982).

\bibitem{Khairutdinov1}
R.F. Khairutdinov and N. Serpone, Prog. React. Kinetics {\bf 21}, 1-68 (1996).

\bibitem{KKO}
Z. Konkoli, A. Karlsson, and O. Orwar, J. Phys. Chem. B {\bf 107}, 14077
  (2003).

\bibitem{Akingbehin}
K. Akingbehin, BioSystems {\bf 35}, 223 (1995).

\bibitem{AkinCon}
K. Akingbehin and M. Conrad, Journal of Parallel and Distributed Computing {\bf
  6}, 245 (1989).

\bibitem{Conrad4}
M. Conrad, European Journal of Operational Research {\bf 30}, 280 (1987).

\bibitem{KirCon1}
K.G. Kirby and M. Conrad, Physica D{\bf 22}, 205 (1986).

\bibitem{KirCon2}
K.G. Kirby and M. Conrad, Bulletin of Mathematical Biology {\bf 46}(5/6), 765
  (1984).

\bibitem{KamCon1}
R. Kampfner and M. Conrad, Bulletin of Mathematical Biology {\bf 45}(6), 931
  (1983).

\bibitem{KamCon2}
R. Kampfner and M. Conrad, Bulletin of Mathematical Biology {\bf 45}(6), 969
  (1983).

\bibitem{Dittrich}
P. Dittrich, J. Ziegler, and W. Banzhaf, Artificial Life {\bf 7}, 225-275
  (2001).

\bibitem{Hess}
B. Hess and A. Mikhailov, Science {\bf 264}, 223 (1994).

\bibitem{Gillespie1}
D.T. Gillespie, J. Comp. Phys. {\bf 22}, 403 (1976).

\bibitem{Gillespie2}
D.T. Gillespie, J. Phys. Chem. {\bf 81}, 2340 (1977).

\end{thebibliography}

\end{document}